\documentstyle[12pt]{article}

\newcommand{\ba}{\begin{eqnarray}}
\newcommand{\ea}{\end{eqnarray}}
\begin{document}
\pagestyle{plain}

\title{Electromagnetic transition form factors of negative parity 
nucleon resonances 
\thanks{Partially supported by EC-contract number ERB FMRX-CT96-0008}}
\author{M.~Aiello, M.M.~Giannini and E.~Santopinto\\
\and
\begin{tabular}{rl}
      &Dipartimento di Fisica dell'Universit\`a di Genova,\\
      &I.N.F.N., Sezione di Genova\\
      &via Dodecaneso 33, 16164 Genova, Italy\\
      &e-mail:giannini@genova.infn.it\\
\end{tabular}}
\date{}
\maketitle
\noindent
\vspace{4pt}
\begin{abstract}
We have calculated the transition form factors for the electromagnetic 
excitation of the negative parity resonances of the nucleon using 
different models previously proposed and we discuss their results and 
limits  by comparison with experimental data.
\end{abstract}
\begin{center}
\end{center}

\newpage
\noindent
\section{Introduction}

Various Constituent Quark Models have been proposed for the internal structure 
of baryons \cite{fai, fey, is, ci, gi, bil, pl, ven, fdr, ple}. A common 
characteristic is that, although the models use different ingredients, they 
are able to give a satisfactory description of the baryon spectrum and, in 
general, of the nucleon static properties.

The problem is that, in any case, the study of hadron spectroscopy is not 
sufficient to distinguish among the various forms of quark dynamics, that 
is among the various models, and so other observables, such as the 
electromagnetic transition form factors and the strong decay amplitudes, are 
important in testing the models for the internal structure of hadrons. 

In order to perform a systematic study of baryon properties it is useful to 
have some general framework, which allows to formulate or reformulate
the various models and 
compare their results in a consistent way. To this end it has been recently 
shown \cite{pl} that a hypercentral approach to quark dynamics can be used. 
This method is sufficiently general to investigate new dynamical features, 
such as three-body mechanisms, and also to reformulate and/or include the 
currently used two-body potential models. 

Within this framework we have performed a study of the helicity 
amplitudes for the photoexcitation of the nucleon resonances \cite{aie}, 
making use of various types of potentials determined in Ref. \cite{pl}. 
The resulting description of the 
helicity amplitudes is qualitatively good and comparable with the one coming 
from various models proposed in the literature, including those which take 
into account some relativistic kinematic corrections \cite{aie}.

In this paper we present the electromagnetic transition form factors 
calculated within the general framework of \cite{pl} using various 
potentials. The $Q^2$-behaviour is more sensitive to the quark wave functions, 
therefore a detailed analysis of the theoretical inputs becomes 
possible, and moreover the electromagnetic transition form factors are going 
to be accurately measured at TJNAF(CEBAF) \cite{ceb}.

We study the 
excitation of the negative parity resonances. Their energies are really well 
described and therefore any discrepancy with the experimental  data cannot 
be ascribed to a deficiency in the description of the spectrum, but eventually  
to some mechanism which is not present in current approaches, such as 
dynamic and relativistic corrections.

In Sect. 2 we briefly describe the model.  In Sect. 3 the electromagnetic 
transition form factors are evaluated and compared with the data and some 
discussion on the limits of the present non relativistic description is 
also given.  A brief conclusion is given in Sect. 4.

\noindent
\section{The model}
We briefly remind the theoretical framework proposed in \cite{pl}, which is a 
three-body force approach to the non-relativistic constituent quark model. 
The internal quark motion is described by the Jacobi coordinates 
$\vec{\rho}$ and $\vec{\lambda}$:
\ba
\vec{\rho}~=~ \frac{1}{\sqrt{2}}
(\vec{r}_1 - \vec{r}_2) ~,\nonumber\\
\\
~~~~\vec{\lambda}~=~\frac{1}{\sqrt{6}}
(\vec{r}_1 + \vec{r}_2 - 2\vec{r}_3) ~\nonumber
\ea
or equivalently, $\rho$, $\Omega_{\rho}$, $\lambda$, $\Omega_{\lambda}$. In 
order to describe three-quark dynamics it is convenient to introduce the 
hyperspherical coordinates, which are obtained substituting the absolute 
values $\rho$ and $\lambda$ by 
\begin{equation}
x=\sqrt{{\vec{\rho}}^2+{\vec{\lambda}}^2} ~~,~~ \quad
\xi=arctg(\frac{{\rho}}{{\lambda}}),
\end{equation}
where $x$ is the hyperradius and $\xi$ the hyperangle. In this way one can use 
the hyperspherical harmonic formalism \cite{mo, fa, hh}. 

The quark potential,
$V$, is assumed to depend on the hyperradius $x$ only, that is to be 
hypercentral. Therefore, $V~=~V(x)$ is in general a three-body potential, 
since the 
hyperradius $x$ depends on the coordinates of all the three quarks. This 
class of potentials
contains also contributions from two-body potentials in hypercentral
approximation \cite{hca, jmr}. For hypercentral potentials, the Schr\"{o}dinger 
equation, in the hyperspherical coordinates, is simply reduced to a single 
hyperradial equation, while the angular and hyperangular parts of the 
3q-states are the known hyperspherical harmonics \cite{fa}. 

There are at least two hypercentral potentials which can be solved analytically. One can 
observe that 
the h.o. potential, which has a two-body character, turns out to be exactly 
hypercentral, since
\begin{equation}
\sum_{i<j}~\frac{1}{2}~k~(\vec{r_i} - \vec{r_j})^2~=~\frac{3}{2}~k~x^2~=
~V_{h.o}(x)
\end{equation}
The other one is the 'hypercoulomb' potential \cite{ferr, nc, hca, br}
\begin{equation}
V_{hyc}(x)= -\frac{\tau}{x}.
\end{equation}\label{hcb}
This potential is not confining, however it has interesting properties. It
 leads to a power-law behaviour of the 
proton form factor \cite{ferr, nc} and of all the transition form factors 
\cite{sig} and has a perfect degeneracy between 
the first $0^+$ excitated state and the first $1^-$ states \cite{lip, ferr,
rpl, br}, which can be respectively identified with the Roper resonance and 
the negative parity resonances. This degeneracy seems to be in agreement with
phenomenology and is typical of an underlying O(7) symmetry \cite{br}.

Besides the two analytical solutions, we have studied three-body potentials of 
the form \cite{pl}
\begin{equation}\label{eq:pot}
V(x)= -\frac{\tau}{x}~+~\frac{\kappa}{x^2}~+~\beta x.
\end{equation}
The hypercentral equation is solved numerically. Starting 
from any potential $V(x)$ and solving the corresponding hyperradial equation, 
one can construct a complete basis of antisymmetric three-quark states, 
analogously to what is done in standard  h.o. models \cite{is}, combining the 
$SU(6)$-spin-flavour configurations with the space wave functions \cite{pl}. 
In order to account for the splitting within each $SU(6)$-multiplet, 
a hyperfine interaction has been added and treated as a perturbation and 
therefore each resonance is a superposition of $SU(6)$-configurations.
 In this way, 
one obtains \cite{pl} a description of the observed spectrum for different 
choices of the potential parameters of Eq. \ref{eq:pot}.

\noindent
\section{Electromagnetic transition form factors}

The electromagnetic transition form factors, 
$A_{1/2}(Q^2)$ and $A_{3/2}(Q^2)$, are defined as the transition matrix elements of the 
transverse electromagnetic interaction, $H_{e.m.}^t$, between the nucleon, 
$N$, and the resonance, $B$, states:
\begin{equation}
\begin{array}{rcl}
A_{1/2}(Q^2)&=& \langle B, J', J'_{z}=\frac{1}{2}\ | H_{em}| N, J~=~
\frac{1}{2}, J_{z}= -\frac{1}{2}\
\rangle\\
& & \\
A_{3/2}(Q^2)&=& \langle B, J', J'_{z}=\frac{3}{2}\ | H_{em}| N, J~=~
\frac{1}{2}, J_{z}= \frac{1}{2}\
\rangle\\\end{array}
\end{equation}
The transition operator is assumed to be
\begin{equation}\label{eq:htm}
H^t_{em}~=~-~\sum_{i=1}^3~\left[\frac{e_j}{2m_j}~(\vec{p_j} \cdot \vec{A_j}~+
~\vec{A_j} \cdot \vec{p_j})~+~2 \mu_j~\vec{s_j} \cdot (\vec{\nabla} 
\times \vec{A_j})\right]~~,
\end{equation}
where spin-orbit and higher order corrections are neglected 
\cite{cko, ki, aie}. In 
Eq. \ref{eq:htm} $~~m_j$, $e_j$, $\vec{s_j}$ , $\vec{p_j}$ and $\mu_j~=~\frac{ge_j}{2m_j}$
denote the mass, the electric charge, the spin, the momentum and the magnetic 
moment of the j-th quark, respectively, and $\vec{A_j}~=~\vec{A_j}(\vec{r_j})$
is the photon field. For the transverse interaction it is sufficient, without 
loss of generality, to consider photons with right-handed polarization 
($\epsilon_+~=~-\frac{1}{\sqrt{2}}~(1,i,0)$) and momentum along the z-axis. 
Taking into account the antisymmetry of states, which permits to write 
$H_{e.m.}^t~=~3H_{e.m.}^t(3)$, the transverse coupling is given by
\begin{equation}
H^t_{em}~=~6\sqrt{\frac{\pi}{k_0}}\mu_p~e(3)~e^{ikz_3}~\left[~k(s_{3,x}~+~i
s_{3,y})~+~\frac{1}{g}(p_{3,x}~+~ip_{3,y})\right]~~,
\end{equation}
where $\mu_p$ is the proton magnetic moment and $(k_0,\vec{k})$ is the 
virtual photon tetramomentum. We calculate the transition form factors in 
the Breit frame, using the relation
\begin{equation}
{\vec{k}}^2 = Q^2 + \frac{(W^2 - M^2)^2}{2(M^2 + W^2) + Q^2}~,
\end{equation}
where $M$ is the nucleon mass, $W$ is the mass of the resonance and $Q^2~=
~{\vec{k}}^2- k_0^2$. The matrix elements of the e.m. transition operator between 
any two 3q-states are expressed in terms of integrals involving the 
hyperradial wavefunctions and are calculated numerically. The computer code 
has been tested by comparison with the analytical results obtained with the 
h.o. model of Refs. \cite{cko, ki} and with the analytical model of Ref. 
\cite{sig}. 

The calculations are performed using various models:

1) the potential of Eq. \ref{eq:pot} retaining only the hypercoulomb 
and the linear confinement terms \cite{pl} 
with the parameters, $\tau~=~4.59$ and $\beta~=~1.61~fm^{-2}$,  
fitted to the spectrum, plus a standard hyperfine interaction
\cite{is};

2) the analytical model of \cite{sig}, which corresponds to the 
potential of Eq. 5 with $\tau~=~6.39$ and $\beta~=~0.15~fm^{-2}$ plus 
 a hyperfine interaction with a smooth x-dependence;
the spin-spin interaction is assumed to be 
\ba
V^{S}(x)~=~A~e^{-a x}~
\sum_{i<j}~\vec{\sigma_{i}}\cdot\vec{\sigma_{j}}~=\nonumber\\
=~A~e^{-a x}~[2~S^2-\frac{9}{4}]~~~~,
\ea
where S is the total spin of the 3-quark system,
and the tensor interaction
\ba
V^{T}(x)~=~B~\frac{1}{x^3}~
\sum_{i<j}~\left[\frac{\left(\vec{\sigma_{i}} \cdot
(\vec{r_{i}}-\vec{r_{j}})\right)
~\left(\vec{\sigma_{j}}\cdot(\vec{r_{i}}-\vec{r_{j}})\right)}
{|\vec{r_{i}}-\vec{r_{j}}|^{2}}
-\frac{1}{3}(\vec{\sigma_{i}}\cdot\vec{\sigma_{j}})\right]~,
\label{tens}
\ea
with $A~=~140.7~MeV$, a$~=~1.53 ~fm^{-1}$ and $B~=~14~MeV~fm^3$.
All these parameters are fixed by the reproduction of the spectrum.

3) the potential of Eq. \ref{eq:pot} with $\tau~=~1.6$ and 
$\kappa~=~-0.875~fm$, 
plus the same hyperfine interaction as in  model 1); 
 although not confining this potential has the property of 
reproducing exactly the dipole fit of the proton form factor \cite{noi};

4) the harmonic oscillator with the parameter  $\alpha~=~0.229~GeV$  
which reproduces the proton charge radius \cite{is};

5) the harmonic oscillator  with the parameter $\alpha~=~0.410~GeV$ 
corresponding to a confinement radius of the order of 
$0.5 ~fm$, required in order to reproduce the 
$A^{p}_{3/2}$  at $Q^2~=~0$ for  the $D_{13}(1520)$-resonance 
\cite{cko,ki}.

We give the results for the transition form factors of the negative parity
resonances, considering the excitations for which there is 
some experimental information, that is the $D_{13}(1520)$, $S_{11}(1535)$, 
$S_{11}(1650)$, $S_{31}(1620)$ and $D_{33}(1700)$. The experimental data 
are in some case available up to $Q^2~=~3~(GeV/c)^2$ \cite{burk}. Although 
the CQMs are 
in principle not applicable to such high $Q^2-$values, nevertheless we 
consider that it could be interesting to show the theoretical calculations 
even for $Q^2$ higher than $1~(GeV/c)^2$ and discuss their limits.

In Fig. 1 we report the proton helicity amplitudes of the 
$D_{13}(1520)$-resonance.  The two potentials 1) and 2) give rise 
to similar values.  They both fit the energy levels and lead to a 
confinement radius of the order of $0.5$ fm. 
The medium $Q^2$-behaviour is good but they fail to reproduce well the 
data at low $Q^2$ especially in the $A^p_{\frac{3}{2}}$ case.

In the h.o. model 5) the $0.5$ fm value for the radius is imposed by hand 
in order to reproduce the experimental value of the $D_{13}$-resonance 
 $A^p_{\frac{3}{2}}$ at $Q^2~=~0$ \cite{cko, ki}.
The results are however very different from potential 1) and 2)
and in the $A^p_{\frac{1}{2}}$ case also far from the data. The 
potential which reproduces exactly the dipole form factor, 3), gives too 
damped results; the 
same happens for the h.o. with the correct proton radius, which 
causes a too strong damping in the wave functions.

Similar conclusions can be drawn from Fig. 2, where we show the results for 
the $S_{11}(1535)$-resonance. The two potentials 1) and 2) give a 
reasonable account of data, while the dipole-fit potential 3) is again too 
low with respect to the experimental values. Here we report only one h.o. 
calculation, the 
one with the smaller confining radius, taking into account the mixing of 
$SU(6)$-configurations for the $S_{11}$ states \cite{ikk}. 

The transition form factor for the $S_{11}(1650)$-resonance (see Fig. 3) is 
non zero because of the configuration mixing. The two potentials 1) and 
2) have different hyperfine interactions, nevertheless their results are still 
quite similar. The h.o. is the same as in Fig. 2, while the 
dipole-fit potential is omitted for the reasons stated above.
In Figs 4, 5 and 6 we give the transition form factors for the $S_{31}(1620)$-
and $D_{33}(1700)$-resonances, respectively. Here again the first two 
potentials give similar transition form factors.

All these 
results seem to favour potentials leading to wave functions which are
localized in a small confinement region, of the order of $0.5~fm$.  On the 
other hand, in this 
approach one can see that the reproduction of the 
elastic form factor is not a guaranty of describing also the transition 
form factors. 

It should be stressed that our main concern is the study of the 
possibilities and limits of quark models in the description of dynamical 
properties at low $Q^2$. From the analysis of our results it is evident
that none of the present CQMs can explain adequately the transition form 
factors at low momentum transfer. In our opinion, this discrepancy 
indicates that some important effect at low momentum transfer is missing, 
such as the polarization effect of the Dirac sea, which is not included in 
CQMs. 

The calculations, at variance with what expected, are 
in agreement with the few existing 
data at $Q^2~=~1-2~(GeV/c)^2$, that is outside the range of applicability 
of a non relativistic description.

The problem of a relativistic description is still open. 
The approaches used up to now are mainly of two types. In the first one, 
relativity is taken into account consistently \cite{geig}, but the 
corresponding results are, for the moment, available only for the meson 
spectrum. In the second approach, relativistic corrections are included 
within a light-cone method \cite{ck,sal} or considered as higher order 
contributions to the e.m. current \cite{cl,cp,des}. From these works
one can see that the 
relativistic corrections modify slightly the high-$Q^2$ behaviour, that is 
at $2-3~(GeV/c)^2$, of the form factors \cite{ck,des}. It has been shown that 
significant 
modifications are induced by relativistic corrections at low $Q^2$ 
\cite{ck,des}, 
although the results are still not able to reproduce the data.

\noindent
\section{Conclusions}

We have calculated the transition form factors for the 
excitation of the negative parity nucleon resonances using the general 
framework provided by the three-body force approach of ref. \cite{pl} which
allows to investigate the predictions of new three-body models.

 The h.o. results depend strongly on the 
confinement radius and the $Q^2$-behaviour is not realistic (see also 
Ref. \cite{sw}). 

The dipole-fit potential (model 3)) reproduces exactly the 
elastic proton form factor, however it fails in the case of the 
transition amplitudes. 

From the analysis of our results, one sees that a 
potential containing a hypercoulomb and a linear confinement term  plus a 
hyperfine term (models 1) and 2) ), is able to 
give a reasonable description of the transition form factor data, 
specially at medium values of the 
momentum transfer $Q^2$, that is $1-2~(GeV/c)^2$.
 The agreement at medium 
$Q^2$ is not expected to be modified by 
the inclusion of relativistic corrections, since, according to the 
discussion in the previous Section, in this range the relativistic 
corrections are expected to be not so important \cite{ck,des}.

We have observed that the potentials 1) and 2) still have 
problems for low $Q^2$-values. This can be an indication 
that further degrees of freedom, as q\={q}-pairs
\cite{int}, should be included in the CQM in a more explicit way.

\begin{figure}
\vspace{7.cm}
\caption{Comparison between the experimental data for the transition form 
factors $A^p_{3/2}$,$A^p_{1/2}$ for the $D_{13}(1520)$-resonance and the 
calculations with the potentials 1) 
(full curve), 2) (dot-dashed curve), 3) (dashed curve), 4) (the dotted 
curve with the stronger damping) and 5) (the dotted curve with the softer 
damping). The data are from the compilation of Ref. [26] .}
\end{figure}

\begin{figure}
\vspace{7.cm}
\caption{The same as in Fig. 1, for the  $A^p_{1/2}$ of 
the $S_{11}(1535)$-resonance, calculated with the potentials 1), 2), 3) and 5).
The data are from the compilation of Ref. [26] .}
\end{figure}

\begin{figure}
\vspace{7.cm}
\caption{The same as in Fig. 1, for the $A^p_{1/2}$ of 
the $S_{11}(1650)$-resonance, calculated with the potentials 1), 2) and 5).
The data are from the compilation of Ref. [26] .}
\end{figure}

\begin{figure}
\vspace{7.cm}
\caption{The same as in Fig. 1, for the $A^p_{1/2}$ of 
the $S_{31}(1620)$-resonance, calculated with the potentials 1), 2) and 3).
The data are from the compilation of Ref. [26] .}
\end{figure}

\begin{figure}
\vspace{7.cm}
\caption{The same as in Fig. 1, for the $A^p_{1/2}$ of 
the $D_{33}(1700)$-resonance, calculated with the potentials 1), 2), and 3).
The data are from the compilation of Ref. [26] .}
\end{figure}

\begin{figure}
\vspace{7.cm}
\caption{The same as in Fig. 1, for the $A^p_{3/2}$ of 
the $D_{33}(1700)$-resonance, calculated with the potentials 1), 2), and 3).
 The data are from the compilation of Ref. [26] .}
\end{figure}

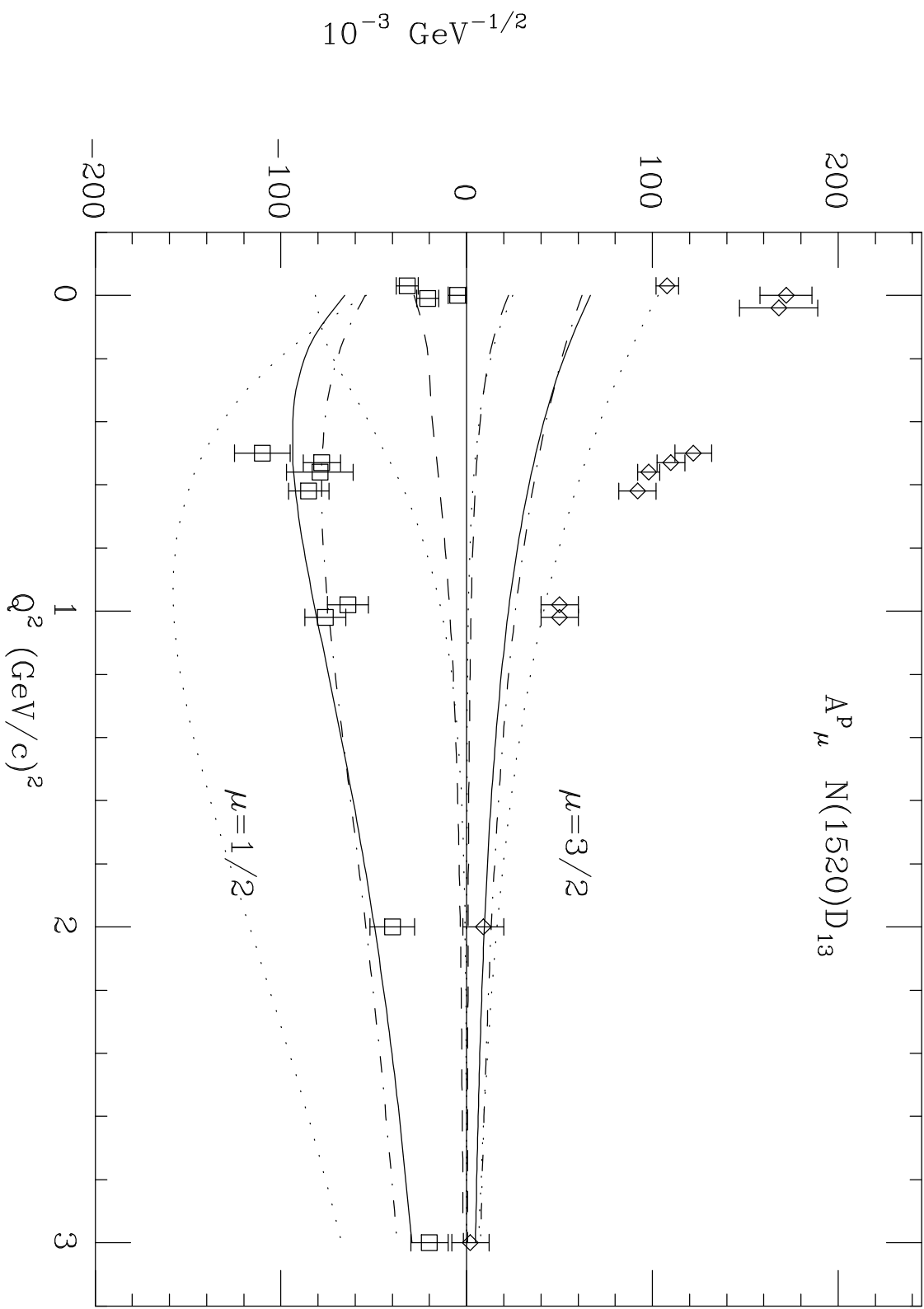
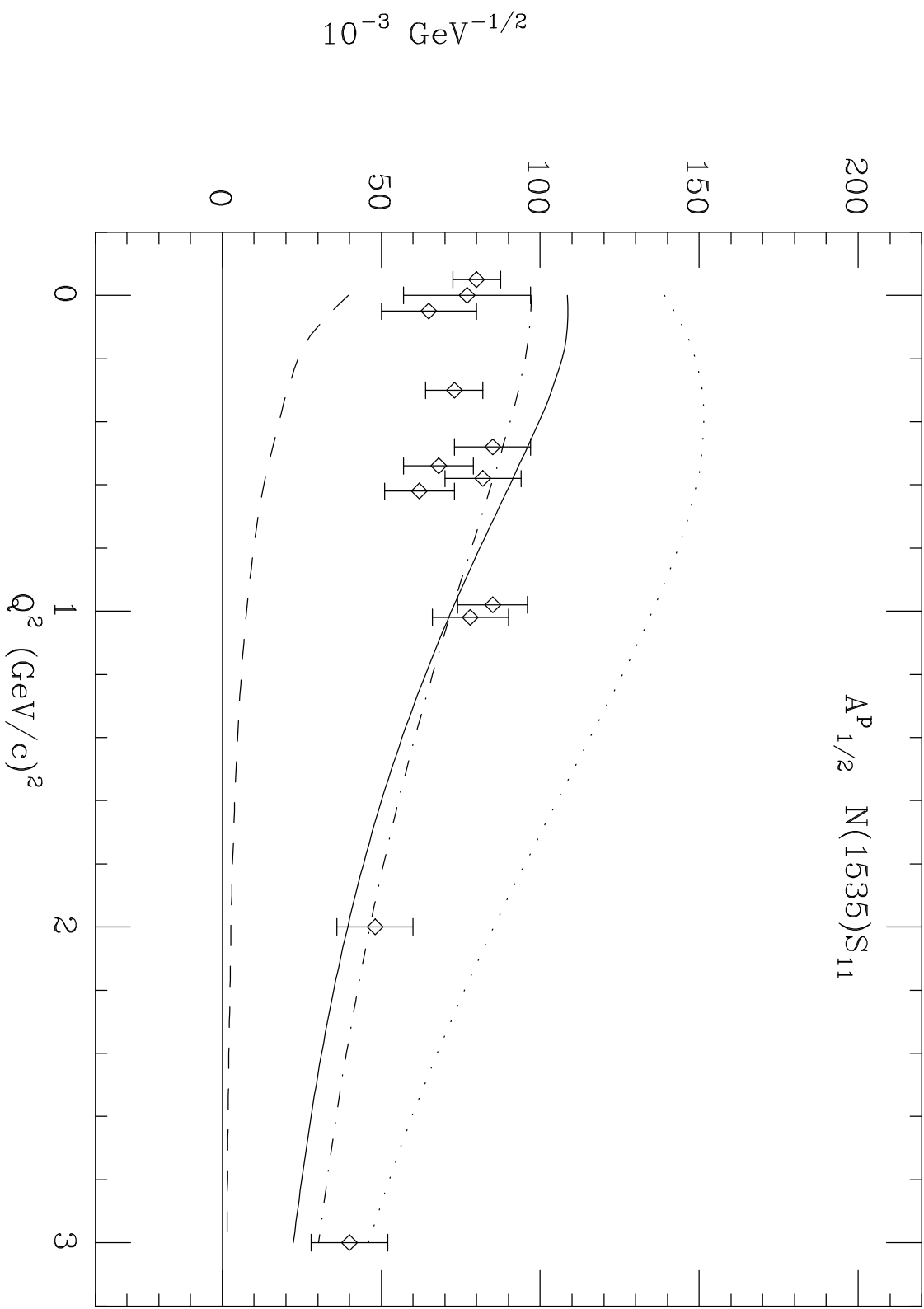
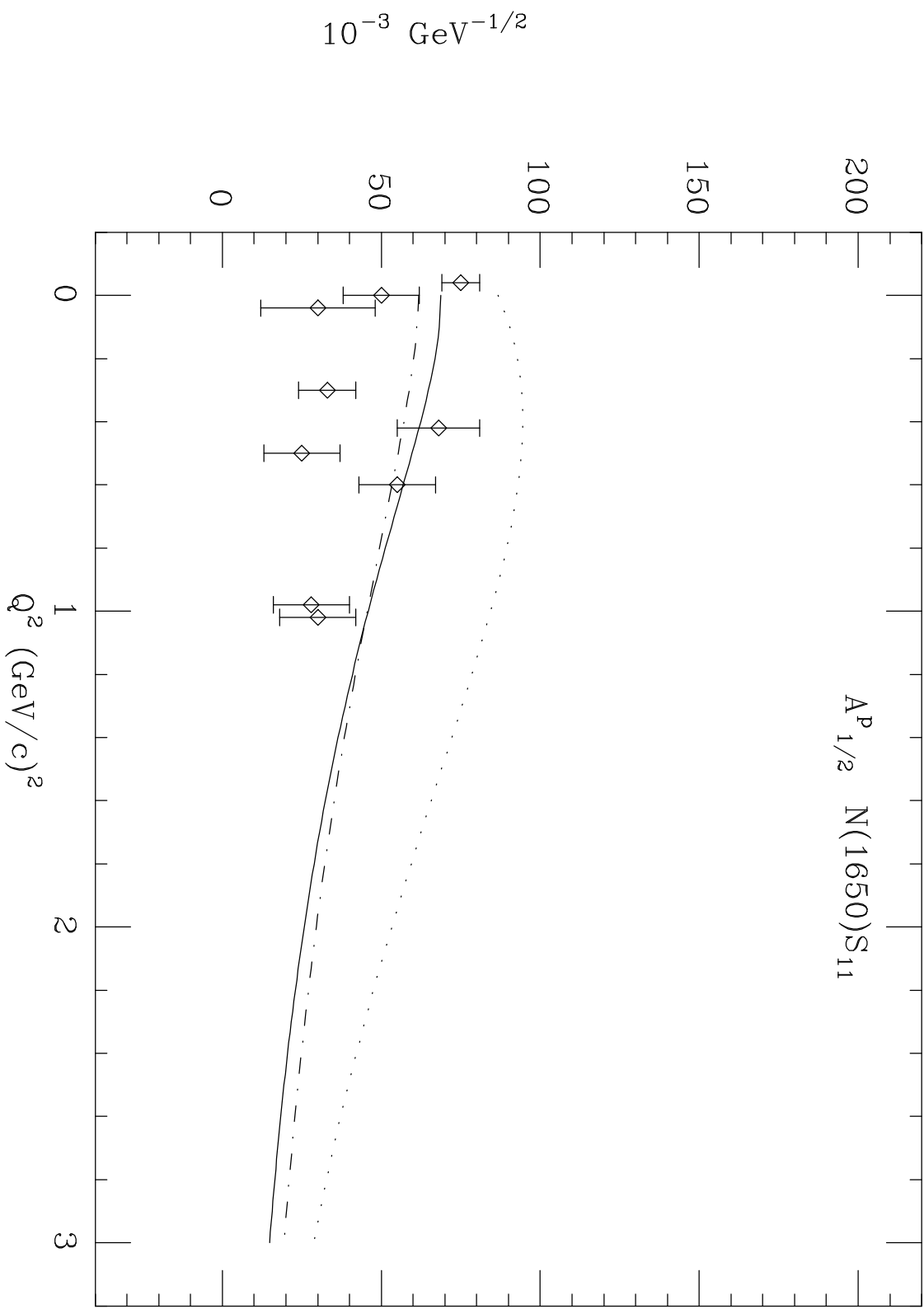
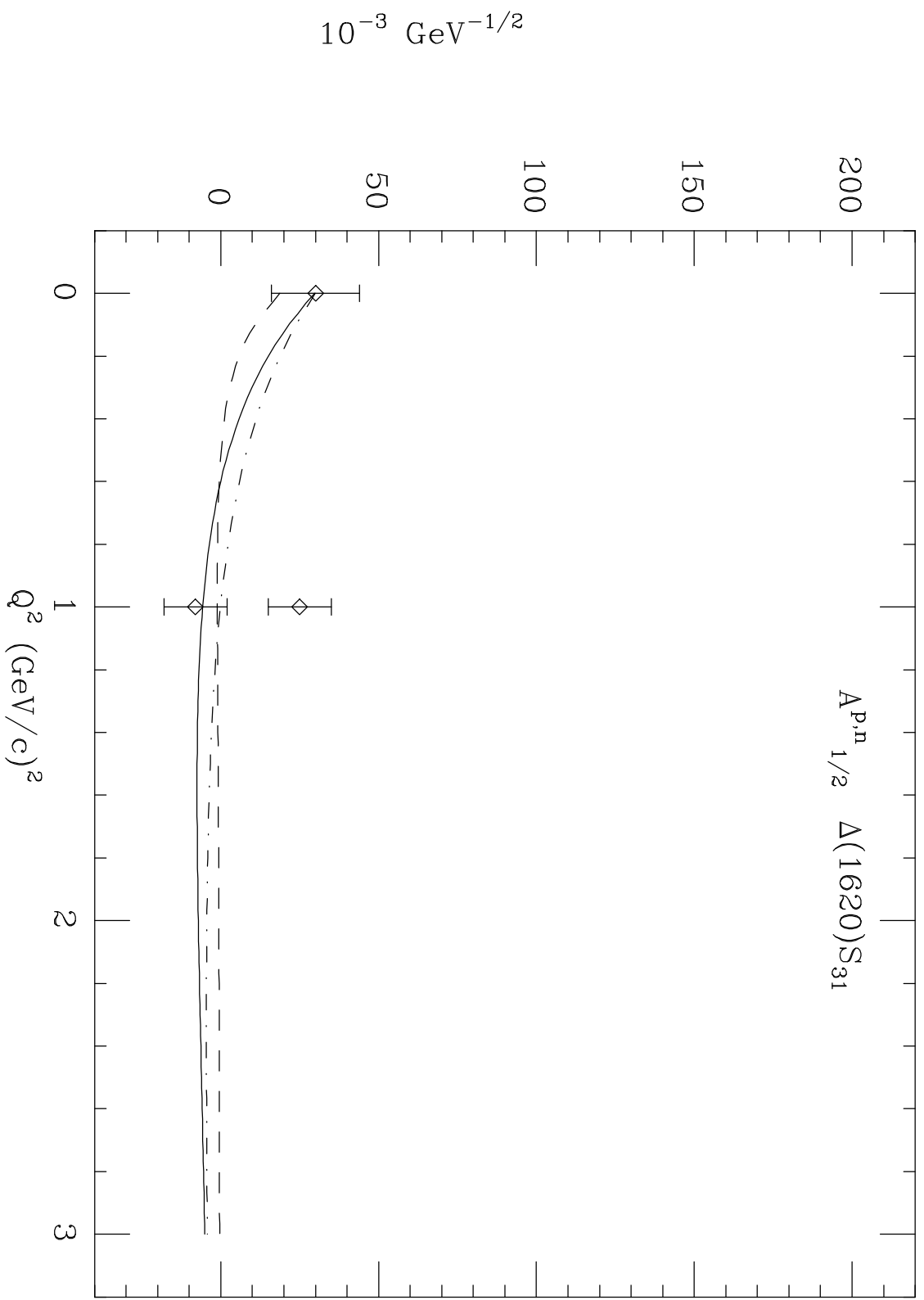
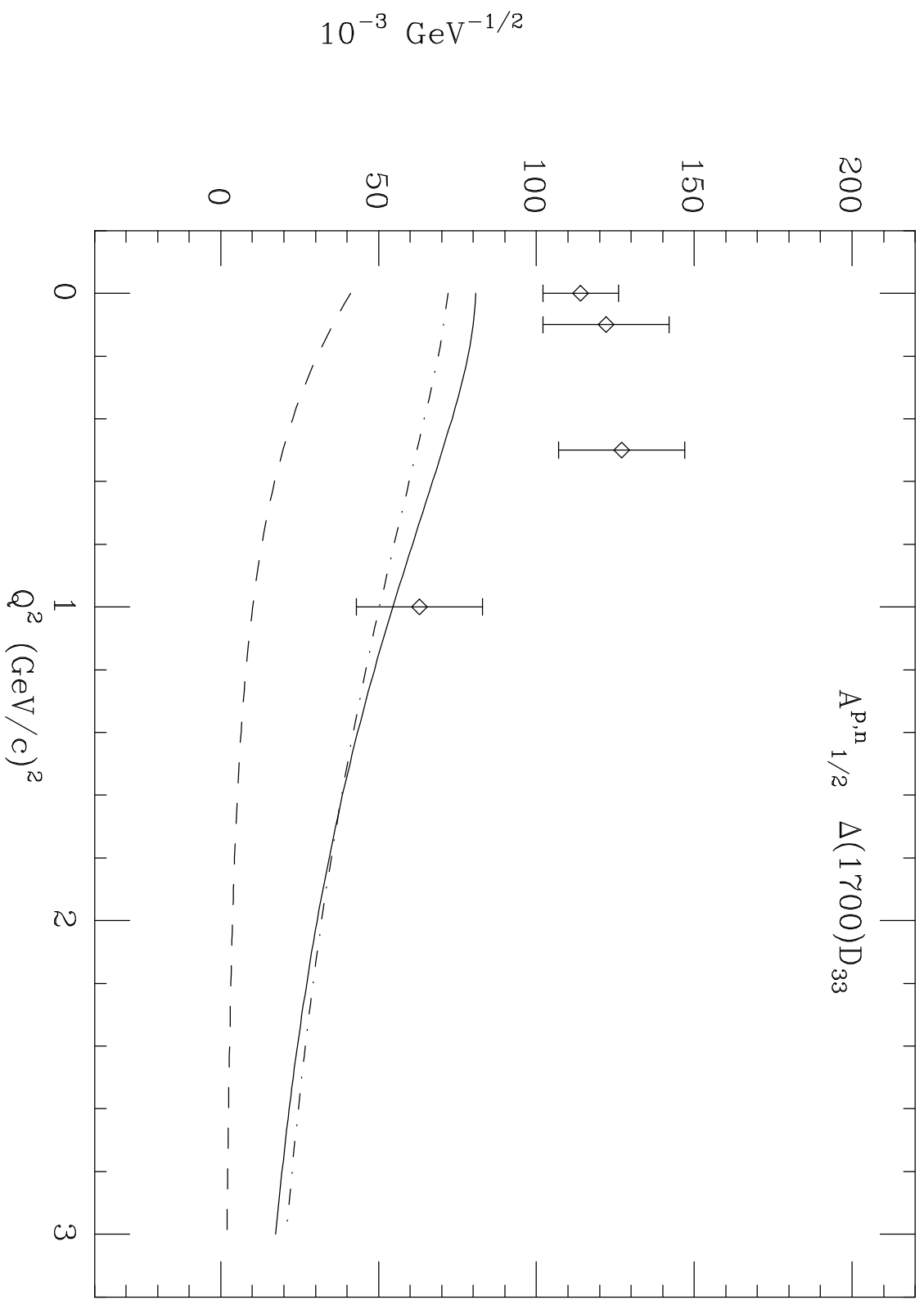
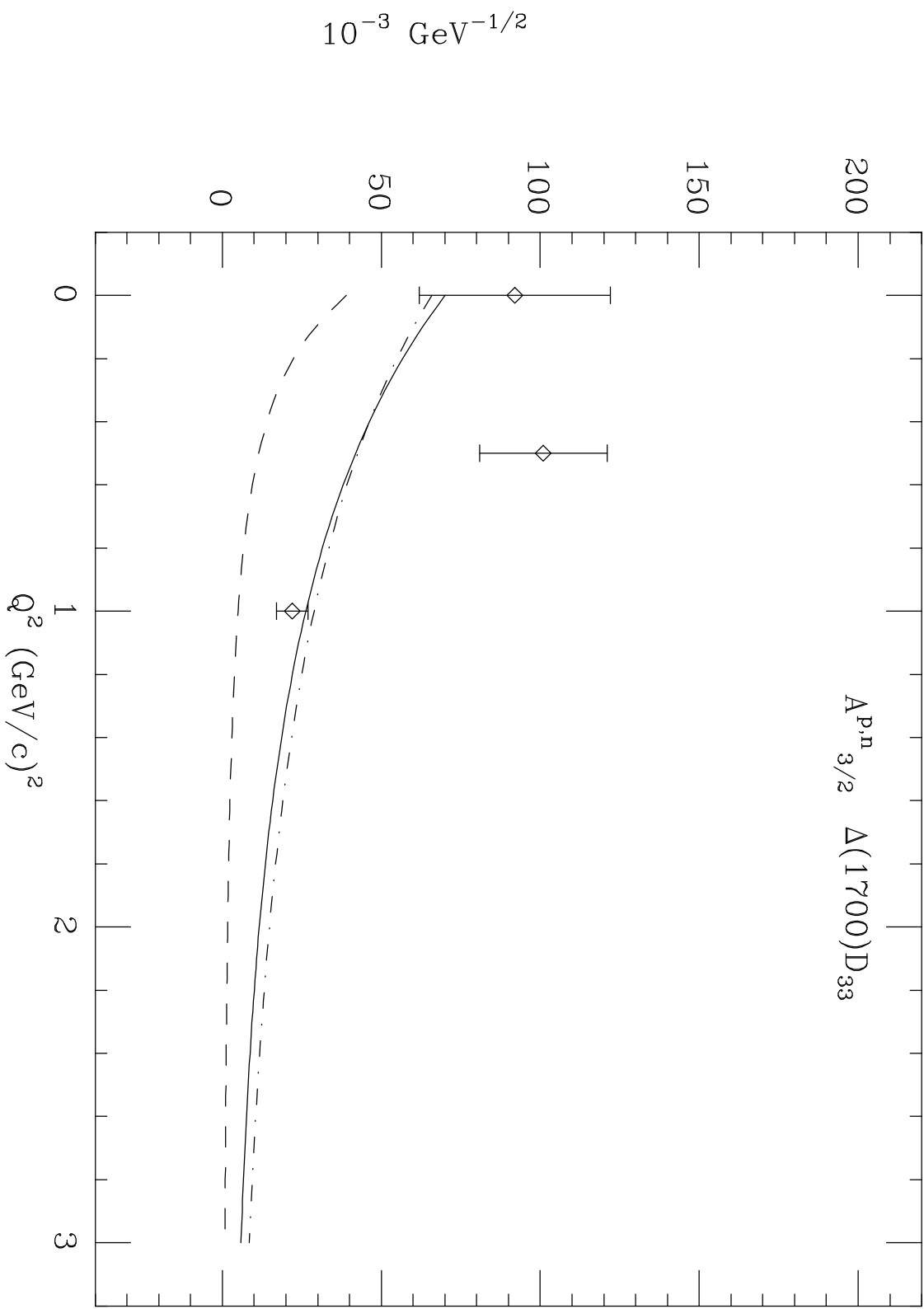
\end{document}